\documentclass[preprint,showpacs,preprintnumbers,amssymb,nofootinbib]{revtex4}

\usepackage{graphicx}

\begin{document}

\preprint{DF/IST-06.2002}

\title{Black hole collision with a scalar particle in 
four, five and seven dimensional anti-de Sitter spacetimes:
ringing and radiation}

\author{Vitor Cardoso}
\email{vcardoso@fisica.ist.utl.pt}
\author{Jos\'e P. S. Lemos}
\email{lemos@kelvin.ist.utl.pt}
\affiliation{
Centro Multidisciplinar de Astrof\'{\i}sica - CENTRA, 
Departamento de F\'{\i}sica, Instituto Superior T\'ecnico,
Av. Rovisco Pais 1, 1049-001 Lisboa, Portugal
}%

\date{\today}

\begin{abstract}
\noindent
In this work we compute the spectra, waveforms and total scalar energy
radiated during the radial infall of a small test particle coupled to
a scalar field into a $d$-dimensional Schwarzschild-anti-de Sitter
black hole. We focus on $d=4,\,5$ and $7$, extending the analysis we
have done for $d=3$. For small black holes, the spectra peaks strongly
at a frequency $\omega \sim d-1$, which is the lowest pure anti-de
Sitter (AdS) mode. The waveform vanishes exponentially as $t
\rightarrow \infty$, and this exponential decay is governed entirely
by the lowest quasinormal frequency.  This collision process is
interesting from the point of view of the dynamics itself in relation
to the possibility of manufacturing black holes at LHC within the
brane world scenario, and from the point of view of the AdS/CFT
conjecture, since the scalar field can represent the string theory
dilaton, and $4$, $5$, $7$ are  dimensions of interest for the
AdS/CFT correspondence.
\end{abstract}


\maketitle

\section{ Introduction}

In this we work we extend the analysis we have done for
3-dimensional anti-de Sitter (AdS) \cite{lemosvitor}, and compute
in detail the collision between a black hole and a scalar particle.
Now, a charged particle following toward a black hole emits radiation
of the corresponding field. Thus, a scalar particle falling into a
black hole emits scalar waves. This collision process is interesting
from the point of view of the dynamics itself in relation to the
possibility of manufacturing black holes at LHC within the brane world
scenario \cite{dim}, and from the point of view of the AdS/CFT
conjecture, since the scalar field can represent the string theory
dilaton, and  $4$, $5$, $7$ (besides $3$) are  dimensions of interest for the
AdS/CFT correspondence \cite{aharonyetal,HH}.  In addition, one can
compare this process with previous works, since there are results for
the quasinormal modes of scalar, and electromagnetic perturbations
which are known to govern the decay of the perturbations, at
intermediate and late times \cite{lemosvitor,HH,cardoso3,cardosolemos1}.

AdS spacetime is the background spacetime in
supersymmetric theories of gravity such as 11-dimensional supergravity
and M-theory (or string theory).  The dimension $d$ of AdS spacetime
is treated as a parameter which, in principle, can have values from
two to eleven in accord to these theories, and where the other spare
dimensions either receive a Kaluza-Klein treatment or are joined as a
compact manifold $\cal M$ into the whole spacetime to yield ${\rm
AdS}_d\times {\cal M}^{11-d}$.  By taking a low energy limits at
strong coupling and by performing a group theoretic analysis,
Maldacena has conjecture a correspondence between the bulk of
$d$-dimenisonal AdS spacetime in string theory and a dual conformal
field gauge theory (CFT) on the spacetime boundary
\cite{maldacenaconjecture}. The first system 
to be studied with care was a D-3brane, in which case the conjecture 
states that type IIB superstring theory in $AdS_5\times S^5$ is the 
same as $\cal N$$=4$ $SU(N)$ super-Yang-Mills (conformal) theory 
on $S^3\times R$, with $\cal N$ being the number of fermionic generators
and $N$ the number of D-branes.
A concrete method to implement this identification was given 
\cite{gubserklebanovpolyakov,witten}, where it was proposed to identify 
the extremum of the classical string theory action $\cal I$ for the 
dilaton field $\phi$, say, at the boundary of AdS, with the generating 
functional $W$ of the Green's correlation functions in the CFT for 
the operator $\cal O$ that corresponds to $\phi$ (in the D-3brane case 
${\cal O}={\rm Tr} F^2$, where $F_{ab}$ is the gauge field strength), 
\begin{equation}
{\cal I}_{\phi_0(x^\mu)}\,=
\nonumber \\
\, W[\phi_0(x^\mu)]\,, 
\label{implement}
\end{equation}
where $\phi_0$ is the value of $\phi$ at the AdS boundary and the
$x^\mu$ label the coordinates of the boundary. The motivation for this
proposal stems from the common substratum of the two AdS/CFT
descriptions, i.e., supergravity theory in the (asymptotically) flat
portion of the full black solutions. Then, a perturbation in this flat
portion disturbs in a similar fashion, both the (soft) boundary in one
description and the CFT on the brane world-volume in the other
description \cite{gubserklebanovpolyakov}.  For systems other than the
D-3brane, analogous statements for the correspondence
AdS$_d$/CFT$_{d-1}$ follow (see the review 
\cite{aharonyetal}). 
In its strongest form the conjecture 
only requires that the spacetime be asymptotically AdS, the interior 
could be full of gravitons or containing a black hole.
The correspondence is indeed a strong/weak duality, and can in
principle be used to study issues of gravity at very strong coupling
(such as, singularities, the localization of the black hole degrees of
freedom and the relation with its entropy, the information paradox,
and other problems) using the associated gauge theory, or CFT issues
such as the difficult to calculate but important n-point correlation
functions using classical gravity in the bulk.  In addition, 
the AdS/CFT correspondence realizes the holographic principle
\cite{thooftsusskind}, since the bulk is effectively encoded in the
boundary.

Some general comments can be made about the mapping AdS/CFT 
when it envolves a black hole. 
A black hole in the bulk corresponds to a thermal 
state in the gauge theory \cite{wittenbanksetal}. Perturbing the 
black hole corresponds to perturbing the thermal state and the 
decaying of the perturbation is equivalent to the return 
to the thermal state. So one obtains a prediction for the 
thermal time scale in the strongly coupled CFT. 
Particles initially far from the black hole correspond to a 
blob (a localized excitation) in the CFT, as the IR/UV 
duality teaches (a  position in the bulk is equivalent 
to size of an object) \cite{susskindwitten}.
The evolution toward the black hole represents a growing 
size of the blob with the blob turning into a bubble travelling 
close to the speed of light \cite{danielsson}.

\newpage

\noindent
\section{The Problem, the Equations and the Laplace transform, and 
the initial and boundary conditions}
\vskip 3mm

\noindent
\subsection{The Problem}
\vskip 3mm
In this paper we shall present the results of the following process:
the radial infall of a small particle coupled to a massless scalar
field, into a $d$-dimensional Schwarzschild-AdS black hole.
We will consider that both the mass $m_0$ and the scalar charge $q_s$ of the
particle are a perturbation on the background spacetime, i.e.,
$m_0\,,q_s\,<< M\,,R$, where $M$ is the mas of the black hole 
and $R$ is the AdS radius. In this approximation the background
metric is not affected by the scalar field, and is given by
\begin{equation}
ds^{2}= f(r) \,dt^{2}- \frac{dr^{2}}{f(r)}-
r^{2}d\Omega_{d-2}^{2}\,,
\label{lineelement}
\end{equation}
where, $f(r)=
(\frac{ r^{2}}{R^2}+1-\frac{16\pi M}{(d-2)A_{d-2}}\frac{1}{r^{d-3}})$,
$A_{d-2}$ is the area of a unit $(d-2)$ sphere, 
$A_{d-2}=\frac{2 \pi^{\frac{d-1}{2}}}{\Gamma(\frac{d-1}{2})}$, and
$d\Omega_{d-2}^{2}$ is the line element on the unit sphere $S^{d-2}$.
The action for the scalar field $\phi$ and particle is given by 
a sum of three parts, the action for the scalar field itself, 
the action for the particle and an interaction piece, 
\begin{equation}
{\cal I}=-\frac{1}{8 \pi} \int \phi _{;a} \phi ^{;a} \sqrt{-g} \,d^dx-
 m_0 \int (1+q_s \phi)(-g_{ab}\dot{z}^a \dot{z}^b)^{\frac{1}{2}} 
d\lambda \,,
\label{action}
\end{equation}
where $g_{ab}$ is the background metric, $g$ its determinant, and 
$z^a(\lambda)$ represents the worldline of the particle as a function
of an affine parameter $\lambda$.

\noindent
\subsection{The Equations and the Laplace transform}
\vskip 3mm
We now specialize to the radial infall case. In the usual
(asymptotically flat) Schwarzschild geometry, one can for example let
a particle fall in from infinity with zero velocity there
\cite{zerillietal}. The peculiar properties of AdS spacetime do not
allow  a particle at rest at infinity \cite{lemosvitor} (we
would need an infinite amount of energy for that) so we consider the
mass $m_0$ to be held at rest at a given distance $r_0$ in
Schwarzschild coordinates. At $t=0$ the particle starts falling into
the black hole.  As the background is spherically symmetric, Laplace's
equation separates into the usual spherical harmonics
$Y(\theta,\varphi_1,..,\varphi_{d-3})$ defined over the $(d-2)$ unit
sphere \cite{bateman}, where $\theta$ is the polar angle and
$\varphi_1,..,\varphi_{d-3}$ are going to be considered azimuthal
angles of the problem. In fact, since we are considering radial
infall, the situation is symmetric with respect to a $(d-3)$
sphere. We can thus decompose the scalar field as
\begin{equation}
\phi(t,r,\theta,\varphi_1,..,\varphi_{d-3})=  \frac{1}{r^{\frac{d-2}{2}}}
\sum_{l}Z_l(t,r) Y_{l0..0}(\theta)\,.
\label{sphericalharmonics1}
\end{equation}
The polar angle $\theta$ carries all the angular information, and $l$ is the
angular quantum number associated with $\theta$. From now on, instead 
of   $Y_{l0..0}(\theta)$, we shall simply write 
$Y_{l}(\theta)$ for the spherical harmonics over the $(d-2)$ unit sphere.
In fact $Y_{l}(\theta)$ is, apart from normalizations, just a Gegenbauer
polynomial $C_{l}^{\frac{d-3}{2}}(\cos\theta)$ \cite{bateman}. 
Upon varying the action (\ref{action}), integrating over the $(d-2)$ sphere
and using the orthonormality properties of the spherical harmonics we
obtain the following equation for $Z_l(t,r)$
\begin{equation}
\frac{\partial^{2} Z_l(t,r)}{\partial r_*^{2}}-
\frac{\partial^{2} Z_l(t,r)}{\partial t^{2}}-
V(r)Z_l(t,r)=
\frac{4\pi q_s m_0 f}{r^{\frac{d-2}{2}}}(\frac{dt}{d\tau})^{-1}
\delta(r-r_p(t))Y_{l}(0) \,.
\label{waveequation1}
\end{equation}
The potential $V(r)$ appearing in equation (\ref{waveequation1}) is given by
\begin{equation}
V(r)=
f(r)\left\lbrack\frac{a}{r^2}+
\frac{(d-2)(d-4)f(r)}{4r^2}+\frac{(d-2)f'(r)}{2r}\right\rbrack \,,
\label{potential}
\end{equation}
where $a=l(l+d-3)$ is the eigenvalue of the Laplacian on $S^{d-2}$,
and the tortoise coordinate $r_*$ is defined as
$\frac{\partial r}{\partial r_*}= f(r)$.
By defining the Laplace transform ${\bf \tilde Z_l}(\omega,r)$ of
$Z_l(t,r)$ as 
\begin{equation}
{\bf \tilde Z}(\omega,r)=
\frac{1}{(2\pi)^{1/2}}\int_{0}^{\infty}e^{i \omega t}Z_l(t,r)dt.
\label{laplace}
\end{equation}   
Then, equation (\ref{waveequation1}) transforms into
\begin{equation}
\frac{\partial^{2} {\bf \tilde Z}(\omega,r)}{\partial r_*^{2}} +
\left\lbrack\omega^2-V(r)\right\rbrack
{\bf \tilde Z}(\omega,r)=S_l(\omega,r)+
\frac{i\omega}{(2\pi)^{1/2}}{Z_0}_l(r) \,,
\label{waveequation2}
\end{equation}
with the source term $S_l(\omega,r)$ given by,
\begin{equation}
S_l(\omega,r)=\frac{2 (2\pi)^{1/2} 
q_s m_0 f Y_{l}(0)}{r^{(d-2)/2}(E^2-f)^{1/2}} 
e^{i \omega T(r)}\,.
\label{source}
\end{equation}
Note that ${Z_0}_l(r)$ is the initial value of $Z(t,r)$, i.e., 
${Z_0}_l(r)=Z(t=0,r)$, satisfying
\begin{equation}
\frac{\partial^{2} {Z_0}_l(r)}{\partial r_*^{2}} 
-V(r){Z_0}_l(r)=
-\frac{4 \pi q_s m_0 f(r)
Y_{l}(0)}{r^{(d-2)/2}}(\frac{dt}{d\tau})_{r_0}^{-1}\delta(r-r_0) \,,
\label{initial}
\end{equation}
where $r_0=r_p(t=0)$. 
We have represented the particle's worldline by
$z^{\mu}=z_p^{\mu}(\tau)$, with $\tau$ the proper
time along a geodesic.
Here, $t=T(r)$ describes the particle's radial trajectory giving the time
as a function of radius along the geodesic
\begin{equation}
\frac{dT(r)}{dr}=-\frac{E}{f(E^2-f)^{1/2}} \,\,
\label{time}
\end{equation}
with initial conditions $T(r_0)=0 $, and $E^2=f(r_0)$.

We have rescaled $r$, $r\rightarrow\frac{r}{R}$, and measure
everything in terms of $R$, i.e., $\omega$ is to be read $\omega R$,
$\Psi$ is to be read $\frac{R}{q_s m_0}\Psi$ and $r_+$, the horizon
radius is to be read $\frac{r_+}{R}$.

\subsection{The Initial Data}
We can obtain ${Z_0}_l(r)$, the initial value of $Z(t,r)$, by
solving numerically equation (\ref{initial}), 
demanding regularity at both the horizon and infinity
(for a similar problem, see for example \cite{wald,burko,ruf}).
To present the initial data and the results we divide the problem 
into two categories: (i) small black holes with $r_+<<1$, and (ii) 
intermediate and large black holes with 
$r_+\buildrel>\over\sim 1$.

\vskip .5cm

{\bf (i): Initial data for small black holes, $r_+=0.1$}

In Fig. 1 we present initial data for small black holes with $r_+=0.1$
in the dimensions of interest ($d=4,\,5$ and $7$).  In this case, the
fall starts at $r_0=0.5$.  Results referring to initial data in $d=3$
(BTZ black hole) are given in \cite{lemosvitor}.
We show a typical form of ${Z_0}_l$ for $r_+=0.1$ and
$r_0=0.5$, and for different values of $l$.  As a test for the
numerical evaluation of ${Z_0}_l$, we have checked that as $r_0
\rightarrow r_+$, all the multipoles fade away, i.e., ${Z_0}_l
\rightarrow 0$, supporting the No Hair Conjecture.
Note that ${Z_0}_l$ has to be small. 
We are plotting $\frac{{Z_0}_l}{q_sm_0/R}$. 
Since $q_sm_0/R<<1$ in our approximation one has from Figs. 1-3 
that indeed ${Z_0}_l<<1$. 
\begin{figure}
\includegraphics{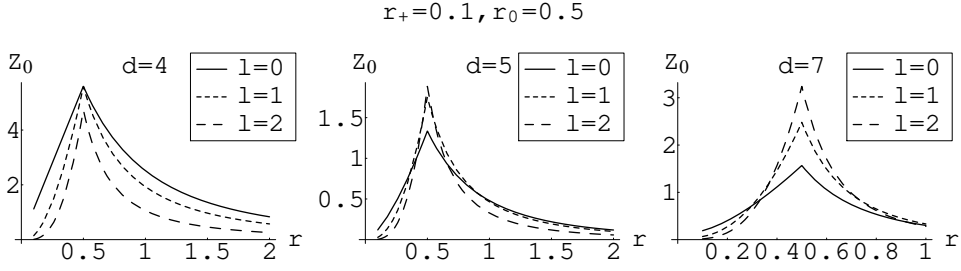}
\caption{\label{static1}
Initial data ${Z_0}_l$ for a small black hole with $r_+=0.1$, for
$d=4,\, 5$ and $7$ (from left to right respectively). The small scalar particle
is located at $r_0=0.5$. The results are 
shown for the lowest values of the angular quantum number $l$. }
\end{figure}
\vskip .5cm

{\bf (ii): Initial data for intermediate and large black holes, $r_+=1$}
\vskip5mm

\begin{figure}
\includegraphics{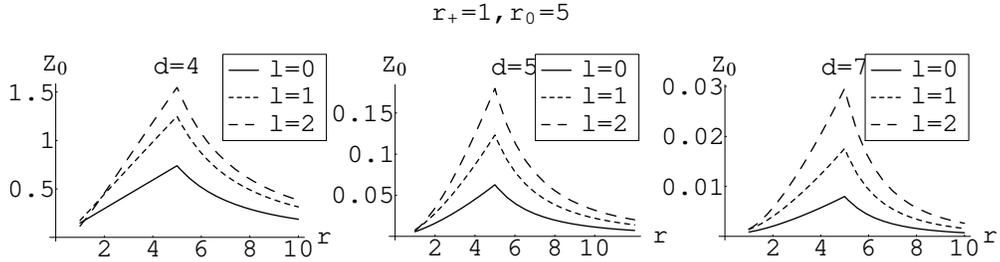}
\caption{\label{static3}Initial data ${Z_0}_l$ for a black hole 
with $r_+=1$, and with the
particle at $r_0=5$, for some values of $l$ the angular quantum
number. Again, we show the results for $d=4,\,5$ and $7$ 
from left to right respectively.  }
\end{figure}
\begin{figure}
\includegraphics{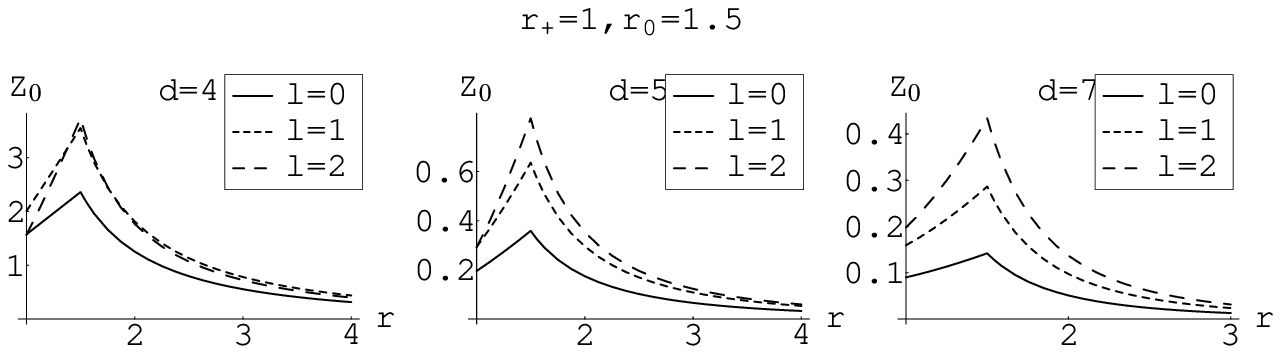}
\caption{\label{static2}Initial data ${Z_0}_l$ for a black hole 
with $r_+=1$, and with the
particle at $r_0=1.5$, for some values of $l$ the angular quantum
number. Again, we show the results for $d=4,\,5$ and $7$ 
from left to right respectively.  }
\end{figure}

In Figs. 2 and 3 we show initial data for an
intermediate to large black hole, $r_+=1$.  In Fig. 2 the fall starts at
$r_0=5$. We show a typical form of ${Z_0}_l$ for $r_+=1$ and
$r_0=5$, and for different values of $l$.  
In Fig. 3 it starts further down at $r_0=1.5$. We 
show a typical form of ${Z_0}_l$ for $r_+=1$ and
$r_0=1.5$, and for different values of $l$.  Again, 
we have checked that as $r_0 \rightarrow r_+$, 
all the multipoles fade away, i.e., ${Z_0}_l
\rightarrow 0$, supporting the No Hair Conjecture.

Two important remarks are in order:
first, it is apparent from Figs. 1-3 that the field (sum over the
multipoles) is divergent at the particle's position $r_0$. This is to
be expected, as the particle is assumed to be point-like;  second, one
is led to believe from Figs. 1-3 (but especially from Figs. 2-3)
that ${Z_0}_l$ increases with $l$.  This is not true however, as this
behavior is only valid for small values of the angular quantum number
$l$. For large $l$, ${Z_0}_l$ decreases with $l$, in such a manner as
to make $\phi(t,r)$ in (\ref{sphericalharmonics1}) convergent and
finite. For example, for $r_+=1, r_0=5$ and $d=4$, we have at $r=6$,
${Z_0}_{l=20}=0.781$ and ${Z_0}_{l=40}=0.3118$.

\subsection{Boundary Conditions and the Green's Function}

Equation (\ref{waveequation2}) is to be solved with the boundary
conditions appropriate to AdS spacetimes, but special attention must
be paid to the initial data \cite{lemosvitor}: ingoing waves at the
horizon,
\begin{equation}
{\bf \tilde Z} \sim F(\omega) e^{-i\omega r_*}+
\frac{i {Z_0}_l}{(2\pi)^{1/2}\omega }\,\,\,, r\rightarrow r_+\,,
\label{boundarybehavior2}
\end{equation}
and since the potential diverges at infinity we 
impose reflective boundary conditions
(${\bf \tilde Z} = 0$) there \cite{avis}.
Naturally, given these boundary conditions, 
all the energy eventually
sinks into the black hole. 
To implement a numerical solution, we note that two independent
solutions ${\bf \tilde Z}^H$ and ${\bf \tilde Z^{\infty}}$ 
of (\ref{waveequation2}),
with the source term set to zero, have the behavior:
\begin{eqnarray}
{\bf \tilde Z}^H \sim e^{-i\omega r_*}\quad\quad r \rightarrow r_+ \,,\\
{\bf \tilde Z}^H \sim A r^{d/2-1} +Br^{-d/2}\quad r \rightarrow \infty\,, \\
{\bf \tilde Z^{\infty}} \sim Ce^{i\omega r_*}+
De^{-i\omega r_*}\quad r \rightarrow r_+ \,,\\
{\bf \tilde Z^{\infty}} \sim r^{-d/2} \quad r \rightarrow \infty \,, \\
\label{behavior}
\end{eqnarray}
Here, 
the wronskian $W$ of these two solutions is a constant, $W=2Ci\omega$. 
Define as in \cite{lemosvitor} $h^H$ through $dh^H/dr_*
=-{\bf \tilde Z}^H$ and $h^{\infty}$ 
through $dh^{\infty}/dr_*=
-{\bf \tilde Z^{\infty}}$.
We can then show that ${\bf \tilde Z}$ given by 
\begin{eqnarray}
&
{\bf \tilde Z}=\frac{1}{W} \left[ {\bf \tilde Z^{\infty}} 
\int _{-\infty}^{r}{\bf \tilde Z}^H S dr_* +
{\bf \tilde Z}^H\int _{r}^{\infty}{\bf \tilde Z^{\infty}} S dr_* \right]+
\frac{i\omega}{(2\pi)^{1/2} W } \left[ {\bf \tilde Z^{\infty}}
\int _{-\infty}^{r}h^H \frac{d{Z_0}_l}{dr_*} dr_* +\right.
\nonumber \\
&
{\bf \tilde Z}^H 
\int _{r}^{\infty}h^{\infty} \frac{d{Z_0}_l}{dr_*} dr_* 
+  
\left. (h^{\infty}{Z_0}_l{\bf \tilde Z}^H -
h^H{Z_0}_l{\bf \tilde Z^{\infty}})(r)
\right]\,, 
\label{Z}
\end{eqnarray}
is a solution to (\ref{waveequation2}) 
and satisfies
the boundary conditions.
In this work, we are interested in computing 
the wavefunction ${\bf \tilde Z}(\omega,r)$ 
near the horizon ($r\rightarrow r_+$). In this 
limit we have
\begin{eqnarray}
&
{\bf \tilde Z}(r\sim r_+)=\frac{1}{W} \left[ {\bf \tilde Z}^H
\int _{r_+}^{\infty}{\bf \tilde Z^{\infty}} S dr_* \right]+&
\nonumber \\
&
\frac{i\omega}{(2\pi)^{1/2}W }
{\bf \tilde Z}^H\left[ \int _{r_+}^{\infty}{\bf \tilde Z^{\infty}} {Z_0}_l dr_*-
(h^{\infty}{Z_0}_l)(r_+) \right]
+\frac{i{Z_0}_l(r_+)}{(2\pi)^{1/2}\omega }\,,&
\label{Z22}
\end{eqnarray}
where an integration by parts has been used.

All we need to do is to find a solution ${\bf \tilde Z_2}$ of the
corresponding homogeneous equation satisfying the above mentioned
boundary conditions (\ref{behavior}), and then numerically integrate
it in (\ref{Z22}).  In the numerical work, we chose to adopt $r$ as
the independent variable, therefore avoiding the numerical inversion
of $r_*(r)$.  To find ${\bf \tilde Z_2}$,the integration (of the
homogeneous form of (\ref{waveequation2})) was started at a large
value of $r=r_i$, which was $r_i=10^5$ typically.  Equation
(\ref{behavior}) was used to infer the boundary conditions ${\bf
\tilde Z_2}(r_i)$ and ${\bf \tilde Z_2}'(r_i)$.  We then integrated
inward from $r=r_i$ in to typically $r=r_++10^{-6}r_+$.  Equation
(\ref{behavior}) was then used to get $C$.

\noindent
\section{Results}
\vskip 3mm

\subsection{Numerical Results}
Our numerical evolution for the field showed that some drastic changes
occur when the size of the black hole varies, so we have chosen to
divide the results in (i) small black holes and (ii) intermediate and
large black holes . We will see that the behavior of these two classes
is indeed strikingly different.  We refer the reader to
\cite{lemosvitor} for the results in $d=3$.

\bigskip

{\bf (i): Wave forms and spectra for small black holes, $r_+=0.1$}

\medskip
We plot the waveforms and the spectra.
Figs. 4-7 are typical plots for small black holes of waveforms 
and spectra for $l=0$ and $l=1$ (for $l=2$ and higher the conclusions 
are not altered).  
They show the first interesting aspect of our
numerical results: for small black holes the $l=0$ signal is clearly
dominated by quasinormal, exponentially decaying, ringing modes with a
frequency $\omega \sim d-1$ (scalar quasinormal frequencies of
Schwarzschild-AdS black holes can be found in
\cite{HH,cardosolemos1}).  This particular limit is a pure AdS mode
\cite{Xpto,konoplya}.  For example, Fig. 4 gives, for $d=4$,
$\omega=2\pi/T \sim 2\pi/(10/4.5)\sim 2.7$.  This yields a value near
the pure AdS mode for $d= 4$, $\omega= 3$. Likewise, Fig. 4 gives 
$\omega \sim 4$ when $d=5$, the pure AdS mode for $d=5$.  All these
features can be more clearly seen in the energy spectra plots, Fig. 6,
where one can observe the intense peak at $\omega \sim d-1$.  The
conclusion is straightforward: spacetimes with small black holes
behave as if the black hole was not there at all. This can be checked
in yet another way by lowering the mass of the black hole. We have
done that, and the results we have obtained show that as one lowers
the mass of the hole, the ringing frequency goes to $\omega \sim 3$
(for $d=4$) and the imaginary part of the frequency, which gives us
the damping scale for the mode, decreases as $r_+$ decreases. In this
limit, the spacetime effectively behaves as a bounding box in which
the modes propagate ``freely'', and are not absorbed by the black
hole.
\begin{figure}
\includegraphics{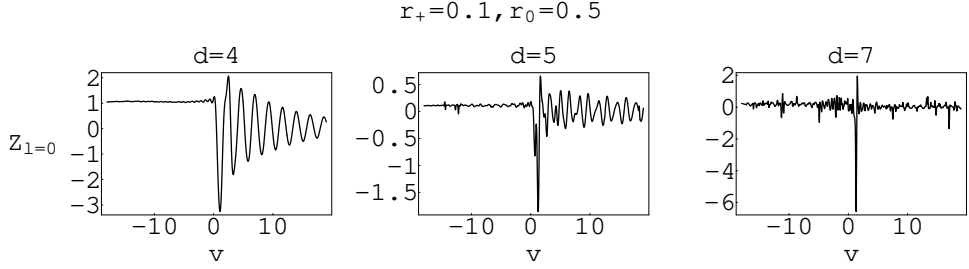}
\caption{\label{waveformsmall1}The spherically symmetric ($l=0$)
waveform for the case of a particle falling from $r_0=0.5$ into a 
$r_+=0.1$ black hole. The results are displayed for $d=4,\,5$ and $7$
from left to right respectively. The coordinate
$v=t+r_*$ is the usual Eddington coordinate.}
\end{figure}
\begin{figure}
\includegraphics{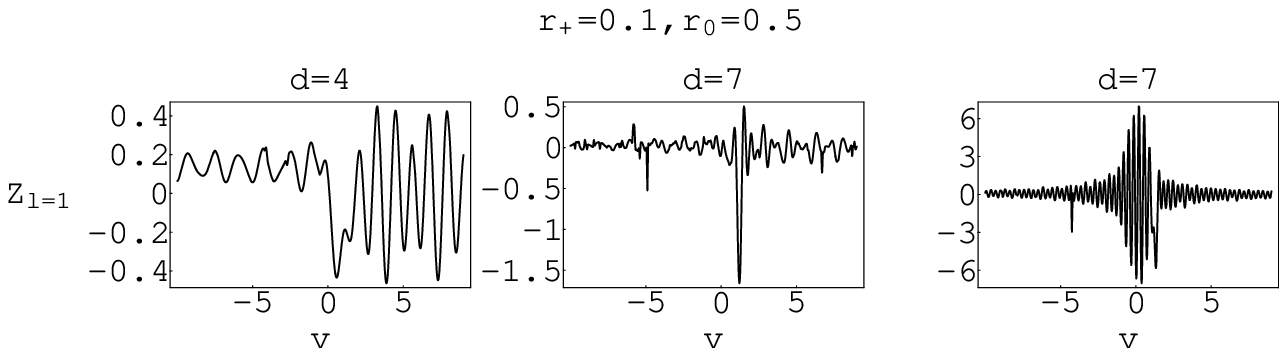}
\caption{\label{waveformsmall2}The $l=1$ waveform for the case of a
particle falling from $r_0=0.5$ into a $r_+=0.1$ black hole.The
results are displayed for $d=4,\,5$ and $7$ from left to right
respectively.}
\end{figure}
\begin{figure}
\includegraphics{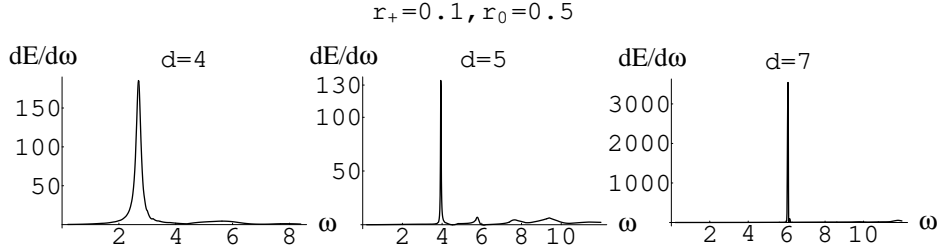}
\caption{\label{spectrasmall1}Typical energy spectra for the spherically
symmetric part of the perturbation ($l=0$), here
shown for $r_+=0.1$, and $r_0=0.5$, and for $d=4,\, 5$ and $7$. 
Total energy in this mode: for d=4 we have $E_{l=0, d=4} \sim 75$. 
For $d=5$, we have 
$E_{l=0, d=5} \sim 34$. For For $d=7$, we have 
$E_{l=0, d=7} \sim 1500$.}
\end{figure}
\begin{figure}
\includegraphics{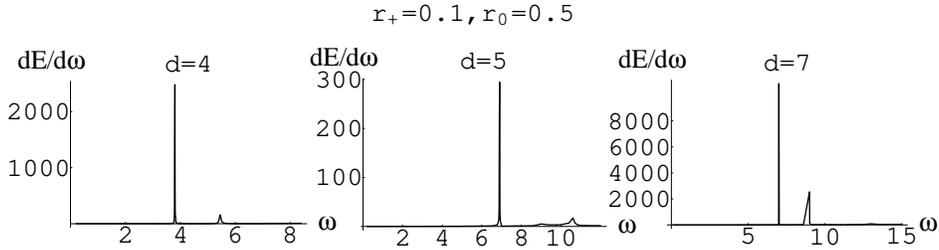}
\caption{\label{spectrasmall2}Typical energy spectra (for $l=1$), here
shown for $r_+=0.1$, and $r_0=0.5$.}
\end{figure}

Not shown is the spectra for higher values of the angular quantum
number $l$. The total energy going down the hole increases slightly
with $l$. This would lead us to believe that an infinite amount of
energy goes down the hole. However, as first noted in \cite{davis2},
this divergence results from treating the incoming object as a point
particle. Taking a minimum size $L$ for the particle implies a cutoff
in $l$ given by $l_{\rm max} \sim \frac{\pi}{2}\frac{r_+}{L}$, and this
problem is solved.

\bigskip

{\bf (ii): Wave forms and spectra for intermediate and large black holes, 
$r_+=1$}
 
\medskip

We plot the waveforms and the spectra.  As
we mentioned, intermediate and 
large black holes (which are of more direct interest to
the AdS/CFT) behave differently. The signal is dominated by a sharp
precursor near $v=r_{*0}$ and there is no ringing: the waveform
quickly settles down to the final zero value in a pure decaying
fashion.  The timescale of this exponential decay is, to high
accuracy, given the inverse of the imaginary part of the quasinormal
frequency for the mode.  The total energy is not a monotonic function
of $r_0$ and still diverges if one naively sums over all the
multipoles.  In either case, there seem to be no power-law tails, as
was expected from the work of Ching et. al. \cite{ching}.  Note that
$E$ is given in terms of $E/(q_sm_0/r)^2$. Since $q_sm_0/r<<1$ the
total energy radiated is small in accord with our approximation.
\begin{figure}
\includegraphics{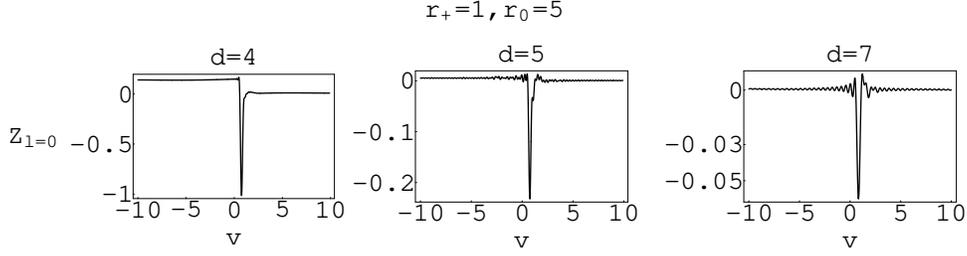}
\caption{\label{waveformlarge1}The spherically symmetric ($l=0$)
waveform for the case of a particle falling from $r_0=5$ into a 
$r_+=1$ black hole. The results are displayed for $d=4,\,5$ and $7$
from left to right respectively.}
\end{figure}
\begin{figure}
\includegraphics{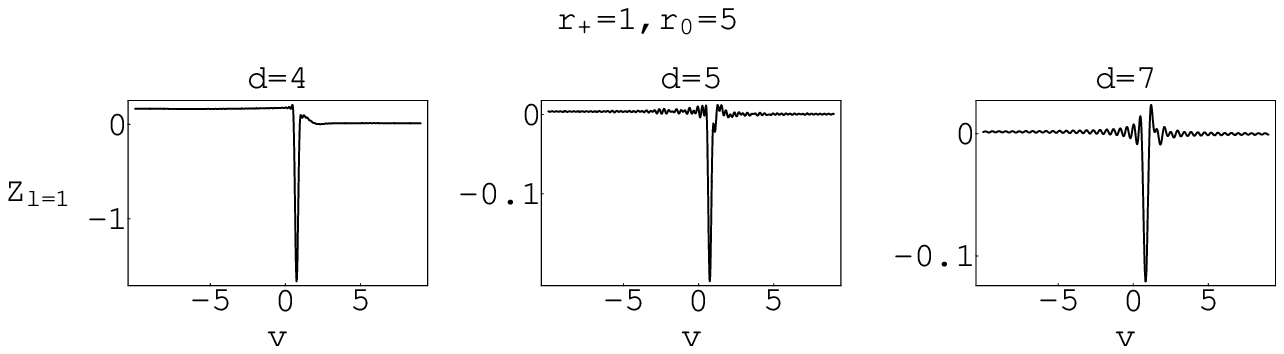}
\caption{\label{waveformlarge2}The $l=1$ waveform for the case of a particle
falling from $r_0=5$ into a $r_+=1$ black hole.}
\end{figure}
\begin{figure}
\includegraphics{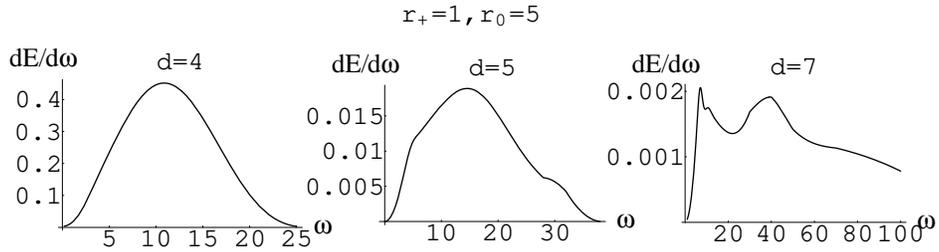}
\caption{\label{spectralarge1}Typical energy spectra for the spherically
symmetric part of the perturbation ($l=0$), here
shown for $r_+=1$, and $r_0=5$, and for $d=4,\, 5$ and $7$.}
\end{figure}
\begin{figure}
\includegraphics{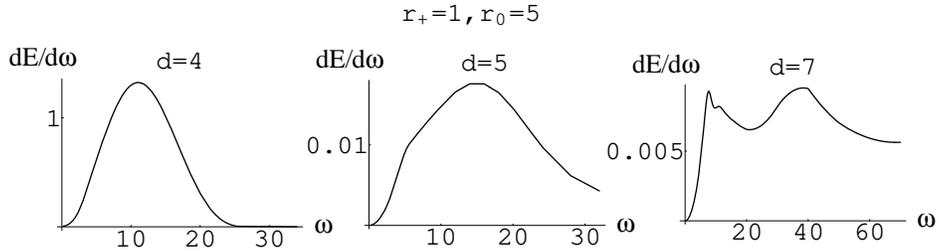}
\caption{\label{spectralarge2}Typical energy spectra (for $l=1$), here
shown for $r_+=1$, and $r_0=5$.}
\end{figure}
The value attained by ${\bf \tilde Z}$ for large negative $v$ -
Fig. 12 - is the initial data, and this can be most easily seen by
looking at the value of $Z_0$ near the horizon in Fig. 3 (see also
\cite{lemosvitor}).  This happens for small black holes also, which is
only natural, since large negative $v$ means very early times, and at
early times one can only see the initial data, since no information
has arrived to tell that the particle has started to fall.  The
spectra in general does not peak at the lowest quasinormal frequency
(cf Figs. 10, 11 and 13), as it did in flat spacetime
\cite{davis}. (Scalar quasinormal frequencies of Schwarzschild-AdS
 black holes can be found in \cite{HH,cardosolemos1}). Most
importantly, the location of the peak seems to have a strong
dependence on $r_0$ (compare Figs. 10 and 13).  This discrepancy has
its roots in the behavior of the quasinormal frequencies.  In fact,
whereas in (asymptotically) flat spacetime the real part of the
frequency is bounded and seems to go to a constant \cite{andersson},
in AdS spacetime it grows without bound as a function of the principal
quantum number $n$ \cite{HH,cardosolemos1}. Increasing the distance
$r_0$ at which the particle begins to fall has the effect of
increasing this effect, so higher modes seem to be excited at larger
distances.

\begin{figure}
\includegraphics{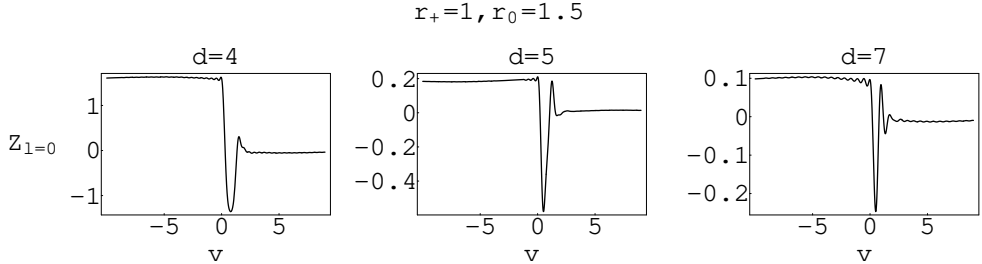}
\caption{\label{waveformlarge3}The $l=0$ waveform for the case of a particle
falling from $r_0=1.5$ into a $r_+=1$ black hole.}
\end{figure}
\begin{figure}
\includegraphics{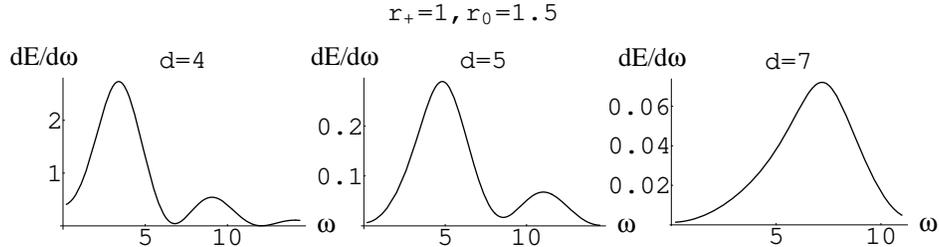}
\caption{\label{spectralarge3} Typical energy spectra (for $l=0$), here
shown for $r_+=1$, and $r_0=1.5$.}
\end{figure}

\noindent
\subsection{Discussion of Results}
\vskip 3mm

Two important remarks regarding these results can be made:

(i) the total energy radiated depends on the size of the infalling object,
and the smaller the object is, the more energy it will radiate.
This is a kind of a scalar analog in AdS space of a well known result
for gravitational radiation in flat space \cite{naka}.

(ii) the fact that the radiation emitted in each multipole is high
even for high multipoles leads us to another important point, first
posed by Horowitz and Hubeny \cite{HH}.  While we are not able to
garantee that the damping time scale stays bounded away from infinity
(as it seems), it is apparent from the numerical data that the damping
time scale increases with increasing $l$. Thus it looks like the late
time behavior of these kind of perturbations will be dominated by the
largest $l$-mode ($L_{\rm max} \sim \frac{r_+}{{\rm 
size}\, {\rm of}\, {\rm object}}$), and
this answers the question posed in \cite{HH}. Thus a
perturbation in $<F^2>$ in the CFT ]
 with given angular dependence $Y_l$ on $S^3$
will decay exponentially with a time scale given by the imaginary part
of the lowest quasinormal mode with {\it that} value of $l$.

\noindent
\section{Conclusions}
\vskip 3mm

We have computed the scalar energy emitted by a point test particle
falling from rest into a Schwarzschild-AdS black hole.
>From the point of view of the AdS/CFT conjecture, where the (large)
black hole corresponds on the CFT side to a thermal state, the
infalling scalar particle corresponds to a specific perturbation of
this state (an expanding bubble), while the scalar radiation is
interpreted as particles decaying into bosons of the associated
operator of the gauge theory.  Previous works \cite{HH,cardosolemos1}
have shown that a general perturbation should have a timescale
directly related to the inverse of the imaginary part of the
quasinormal frequency, which means that the approach to thermal
equilibrium on the CFT should be governed by this timescale. We have
shown through a specific important problem that this is in fact
correct, but that it is not the whole story, since some important
features of the waveforms highly depend on $r_0$.

Overall, we expect to find the same type of features, at least 
qualitatively, in the
gravitational or electromagnetic radiation by test particles falling
into a Schwarzschild-AdS black hole.  For example, if the
black hole is small, we expect to find in the gravitational radiation
spectra a strong peak located at $ \omega^2=4n^2+l(l+1)\;, \quad
n=1,2,...$ \cite{cardoso3}.
Moreover, some major results in perturbation theory and numerical
relativity \cite{anninos,gleiser}, studying the collision of two black
holes, with masses of the same order of magnitude, allow us to infer
that evolving the collision of two black holes in AdS spacetime, should
not bring major differences in relation to our results (though it is of
course a much more difficult task, even in perturbation theory).  In
particular, in the small black hole regime, the spectra and waveforms
should be dominated by quasinormal ringing.


\vskip 1cm

\section*{Acknowledgments} 
This work was partially funded by Funda\c c\~ao para a
Ci\^encia e Tecnologia (FCT) through project PESO/PRO/2000/4014. V. C.
also acknowledges finantial support from FCT through PRAXIS XXI
programme.  J. P. S. L. thanks Observat\'orio Nacional do Rio de
Janeiro for hospitality.

\vskip .5cm


\end{document}